# Transport band gap opening at metal–organic interfaces

Francisc Haidu, Georgeta Salvan, and Dietrich R. T. Zahn

*Semiconductor Physics, Technische Universität Chemnitz, D-09107 Chemnitz, Germany*

Lars Smykalla and Michael Hietschold

*Solid Surfaces Analysis, Technische Universität Chemnitz, D-09107 Chemnitz, Germany*

Martin Knupfer

*Electronic and Optical Properties Department, IFW Dresden, D-01171 Dresden, Germany*

## Abstract

The interface formation between copper phthalocyanine (CuPc) and two representative metal substrates, i.e., Au and Co, was investigated by the combination of ultraviolet photoelectron spectroscopy and inverse photoelectron spectroscopy. The occupied and unoccupied molecular orbitals and thus the transport band gap of CuPc are highly influenced by film thickness, i.e., molecule substrate distance. Due to the image charge potential given by the metallic substrates the transport band gap of CuPc "opens" from (1.4 ± 0.3) eV for 1 nm thickness to (2.2 ± 0.3) eV, and saturates at this value above 10 nm CuPc thickness. The interface dipoles with values of 1.2 eV and 1.0 eV for Au and Co substrates, respectively, predominantly depend on the metal substrate work functions. X-ray photoelectron spectroscopy measurements using synchrotron radiation provide detailed information on the interaction between CuPc and the two metal substrates. While charge transfer from the Au or Co substrate to the Cu metal center is present only at sub-monolayer coverages, the authors observe a net charge transfer from the molecule to the Co substrate for films in the nm range. Consequently, the Fermi level is shifted as in the case of a p-type doping of the molecule. This is, however, a competing phenomenon to the energy band shifts due to the image charge potential.

## I. Introduction

Metal phthalocyanines (MPcs) are chemically and physically stable molecular dyes vastly used in science and technology. Copper phthalocyanine (CuPc) is the most important one and is manufactured on a large scale.[1] It is the most studied molecule in the family of phthalocyanines and porphyrins. Its crystal structure has three polymorphs, among which $\beta$-CuPc is the thermodynamically most stable one.[1]

As underlined by van den Brink and Mopurgo, organic materials are rarely magnetic, however, the MPcs represent an exception to the rule.[2] Recently, CuPc has been considered as a good candidate for molecular spintronics and studies of spin injection and transport properties are reported.[3–5] The most studied hybrid metal–organic (M-O) interfaces are between cobalt, due to its excellent spin filtering properties and different organic molecules, e.g., $Alq_3$ (tris(8-hydroxyquinolato)aluminum),[6,7] CoPc,[5] FePc,[5,8] and especially CuPc.[3–5,8,9]

Ultraviolet photoelectron spectroscopy (UPS) and inverse photoelectron spectroscopy (IPS) are complementary techniques to analyze the occupied and unoccupied electronic states, respectively.[10] CuPc is well characterized in literature by UPS (Refs. 8 and 10–15) and IPS (Refs. 8, 10–12, and 15–18) measurements. Thus, its bulk-like electronic properties were determined on thicker films[10,11,13,16,17] where the influence of the substrate is minimal. Semiconductor–organic (S-O)[12,18] and M-O (Ref. 14) interfaces were characterized by thickness dependent studies starting with ultrathin CuPc coverages in the (sub-)monolayer range up to bulk-like thicknesses (~20 nm). However, this is only half of a typical device structure. In order to study the upper contact interface, metal was evaporated with gradually increasing thicknesses on thicker CuPc films.[8,15] Chemical reactions and interdiffusion[8] were observed. For Co on pentacene, on the other hand, no interdiffusion was detected.[19]

The combination of UPS and IPS techniques was applied to study the impact of the substrate on the electronic properties of CuPc deposited on Au and Co substrates. Previous thickness dependent UPS-IPS studies of CuPc deposited on hydrogen passivated Si(111) reveal a “band bending” behavior,[12] as expected for a S-O interface.[20] At the M-O interface, on the other half, a band gap “opening” is expected, induced by the image charge potential.[20–24] This represents the electric potential produced by the image of the molecular electron cloud mirrored in a conductive, e.g., metal, substrate. Its impact on CuPc molecular films was evidenced in this work by the combined UPS-IPS study. Au, as a nonmagnetic and nonreactive[14] substrate, was employed as reference metal, which is often used as contact in organic electronic devices. The magnetic and chemically reactive Co (Ref. 8) as a substrate for CuPc deposition is of high interest for spintronic applications.

## II. Experiment

In the UPS and additional x-ray photoelectron spectroscopy (XPS) measurements the photoemitted electrons were detected in the direction normal to the sample surface using a hemispherical analyzer. The UPS excitation source was a He discharge lamp using the He I (21.22 eV) excitation line, while for XPS synchrotron radiation was employed. The XPS measurements were performed at the Russian-German beamline at BESSY II, Helmholtz-Zentrum Berlin. The experimental station is equipped with a SPECS PHOIBOS 150 analyzer with a 2D CCD detector. For the acquisition of Cu $2p_{3/2}$ core level spectra we used an excitation energy of 1080 eV, while for N 1s and C 1s the excitation energy was 500 eV. The inverse photoelectron spectrometer is a “home built” system, which operates in the isochromatic mode. It has two main components: a low energy electron gun (electron excitation source) and a Geiger-Müller (GM) tube (photon detector). The collimated beam of the electron gun is monoenergetic, with the energy being spanned by changing the cathode voltage. The current density (in the range of $10^{-6}$ $A/cm^2$, focused on 1 $mm^2$ sample surface) is low enough not to damage the organic molecules. The GM tube is filled with an Ar and ethanol gas mixture and has an $MgF_2$ entrance window. This combination is used as a fixed energy photon detector (10.9 eV). The resolutions of the UPS and IPS setups are 0.20 eV and 0.50 eV, respectively. These were determined from the measurements of the Fermi edge of clean, $Ar^+$ ion sputtered Au and Co foils. All measurements were performed in ultrahigh vacuum conditions with a base pressure in the low $10^{-10}$ mbar range.

The samples were prepared by organic molecular beam deposition thermally subliming CuPc (Sigma-Aldrich® , 99% dye content) from a Knudsen cell onto pure Co (Alfa Aesar, 99.997% Puratronic® ) and Au (MaTecK, 99.995%) foils. The base pressure in the preparation chambers was 5 x $10^{-10}$ mbar. The substrates were cleaned by $Ar^+$ ion bombardment prior to CuPc deposition and the absence of contamination was confirmed by the XP spectra.

The amount of evaporated organic material was monitored by the frequency shift of a quartz crystal microbalance (QCM), which was previously calibrated for transition metal phthalocyanines. The QCM was mounted in the vicinity of the sample. The nominal thicknesses of the investigated CuPc films ranged from 0.5 nm to 22 nm. Films thicker than 20 nm yield no extra information for the thickness dependence study[12,25] and within this thickness range also no charging effects were found.

The spectra (UPS, IPS, and XPS data) analysis was performed with the unifit2010 (Ref. 26) software. For the UPS and IPS data Gaussian peaks and polynomial backgrounds[12] and for the XPS data, Voigt peaks and Shirley backgrounds[26] were employed in the fitting routine.

## III. Results and Discussion

We studied the M-O interface between CuPc and the ferro- and diamagnetic metal substrates of Co and Au, respectively, by performing thickness dependent UPS-IPS measurements. Figure 1 displays the spectra evolution with increasing thickness of the CuPc films on the two metal substrates. From the secondary electron cutoff [Figs. 1(a) and 1(e)], the work functions ($\Phi$) of the substrates and the consecutively deposited CuPc films are determined. The difference between $\Phi_{Co}$ (or $\Phi_{Au}$) and $\Phi_{CuPc}$ defines the interface dipole ($\Delta$).

Figures 1(b) and 1(f) show the evolution of the valence band region with increasing CuPc film thickness on the Co and Au foils, respectively. The spectra plotted by black empty rectangles represent the clean substrates. The UP spectra of Co are smooth and featureless with a well-defined high intensity Fermi edge given by the low lying d-band states. The Au foil on the other hand presents the strong d-band features between 2 eV and 6 eV.

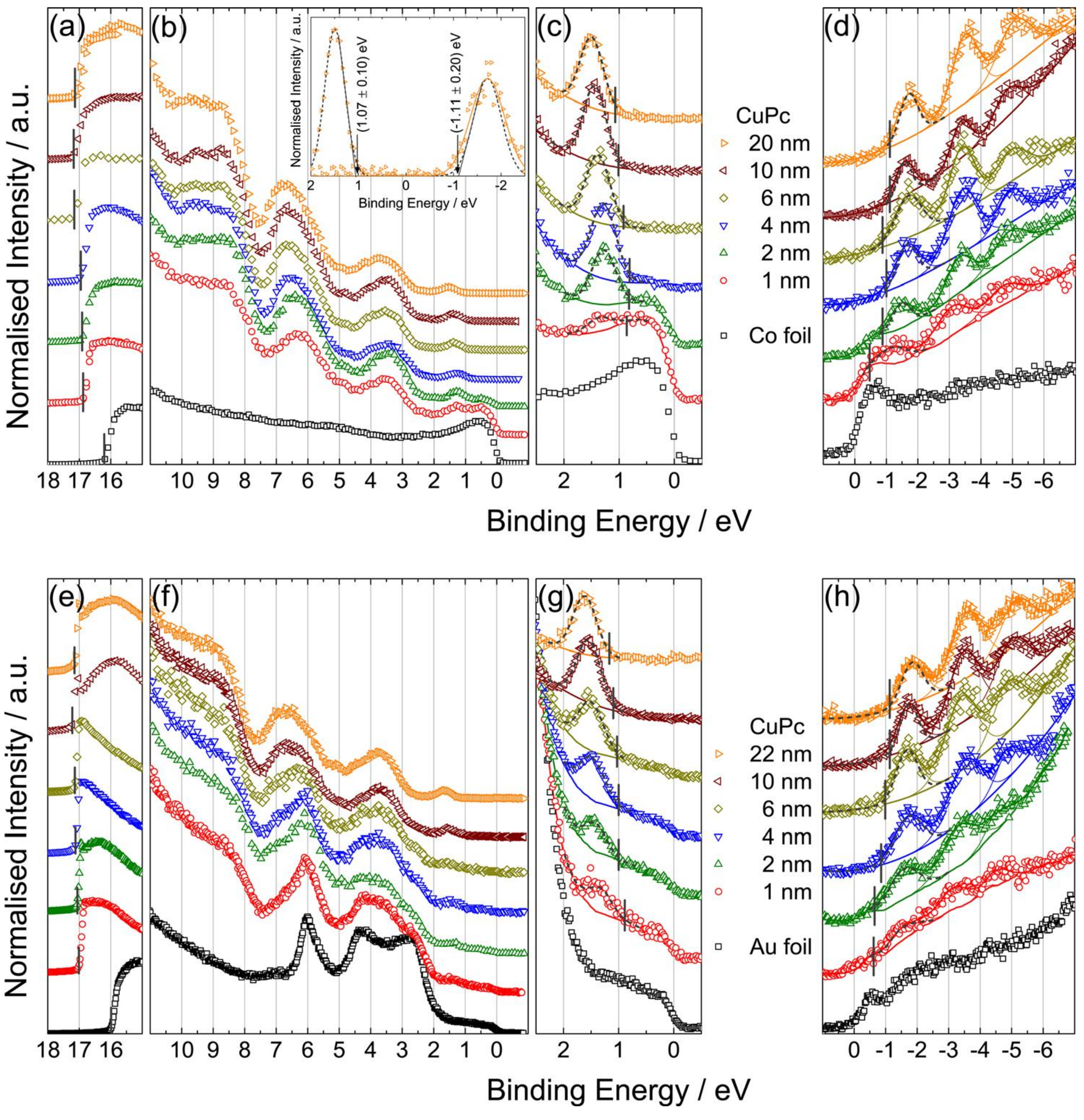

Figure 1. Thickness dependent UPS and IPS measurements on CuPc deposited on Co (a, b, c, d) and on Au foils (e, f, g, h). The secondary electron cutoff (a) and (e), the HOMO onset (c) and (g), and the LUMO onset (d) and (h) positions are marked by vertical bars for CuPc on Co and Au foils, respectively. The valence band region overview is presented in (b) and (f) for CuPc on Co and Au substrates, respectively. The inset in (b) shows how the HOMO and LUMO onset positions were determined for the 20 nm CuPc film.

The features evolving closest to the Fermi edge in Figs. 1(c) and 1(g) represent the highest occupied molecular orbital (HOMO) of CuPc. The HOMO onset positions were determined by fitting the spectra with Gaussian peaks and polynomial backgrounds and are marked in the figure by vertical bars. In case of the thinner films (< 6 nm), the substrate signal was subtracted for data evaluation.

Furthermore, the IP spectra evolution with increasing film thickness is plotted in Figs. 1(d) and 1(h). The Co and Au substrates provide different spectral shapes also in the IPS (conduction band) region: Co has a single feature close to the Fermi edge, while Au has a multitude of features, which belong to different orientations of its polycrystalline surface.[27] The spectra obtained for subsequent CuPc deposition were fitted with three Gaussian peaks. The closest one to the Fermi level represents the lowest unoccupied molecular orbital (LUMO). In order to determine its onset the experimental broadening ($Br_{ex}$ = 0.5 eV) was taken into account and the LUMO peak was deconvoluted[10] according to the formula: $Br_L^2 = Br_{me}^2 - Br_{ex}^2$, where $Br_{me}$ represents the full width at half maximum (FWHM) of the measured peak and $Br_L$ is the FWHM of the peak employed in the LUMO onset determination. The deconvoluted features are plotted in Figs. 1(d) and 1(h) with dashed dark gray lines and their onset positions are marked by vertical bars.

The inset in Fig. 1(b) exemplifies the HOMO and LUMO onset determination for the CuPc (20 nm)/Co sample. The backgrounds of the two spectra were subtracted. Please note that the energy scales in Fig. 1 represent the binding energies (BEs) with regard to the Fermi level ($E_F$ = 0 eV). For continuity of the energy scale all the IP spectra are thus plotted with negative values for the binding energy.

For each film thickness we can identify the following quantities: Φ, Δ, ionization energy (IE), electron affinity (EA), and the transport band gap energy ($E_t$). IE is determined by subtracting the "width" of the UPS signal (measured from the HOMO onset position up to the secondary electrons cutoff) from the photon energy (21.22 eV). In case of thin films, the substrate contribution was subtracted to determine its width. $E_t$ is determined by adding the absolute values of the HOMO and LUMO onset positions. Finally, EA is determined by subtracting $E_t$ from IE. All the determined quantities are plotted in Fig. 2(a) as a function of the CuPc film thickness with regard to the vacuum level ($E_v$ = 0 eV).

First, the work functions of the clean polycrystalline Co and Au substrates were determined to have the values of (5.0 ± 0.1) eV and (5.2 ± 0.1) eV, respectively, in good agreement with literature data.[28] The values are plotted in Fig. 2(a) with full and empty black squares for the Co and Au substrates, respectively, at a nominal CuPc film thickness of 0 nm. The vacuum level positions with respect to $E_F$ are given in Fig. 2(b).

In Fig. 2(a) Φ, IE, EA, Δ, and $E_t$ are plotted as a function of CuPc film thickness. The quantities are given in BE with respect to Ev. Figure 2(b) presents the energy values of $E_v$, the HOMO onset (HOMO), and the LUMO onset (LUMO) with respect to $E_F$. The full and empty symbols in both Figs. 2(a) and 2(b) represent values for measurements on the Co and Au substrates, respectively. Gray symbols are data points for a second set of samples. Each gray symbol represents the quantity given by its colored counterpart and is not explicitly given in the legends. Note that the energy scales are flipped, so that Fig. 2(b) resembled the energy level diagram shown in Fig. 2(c). The dashed, short dashed, and solid lines in Figs. 2(a) and 2(b) are guidelines for the eye. They show the average evolution of the respective quantities with film thickness.

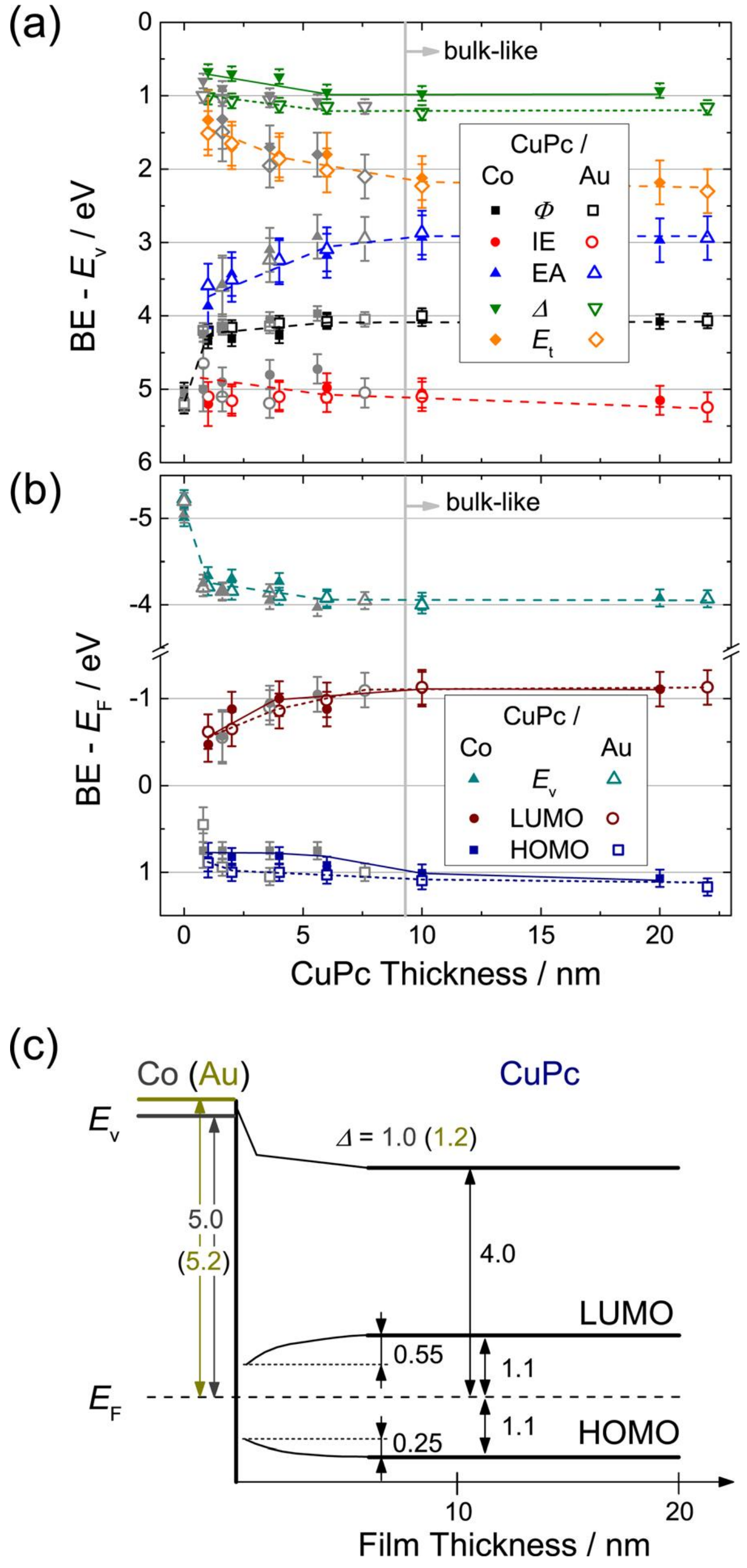

Figure 2. (a) Evolution of the work function (Φ), ionization energy (IE), electron affinity (EA), interface dipole (Δ), and transport band gap ($E_t$) of CuPc films deposited on Co (full symbols) and Au (empty symbols) substrates as a function of film thickness. (b) Energy of the vacuum level ($E_v$), LUMO onset, and HOMO onset positions for CuPc on Co (full symbols) and Au (empty symbols) foils as a function of film thickness. Gray symbols represent the second set of measurements on new samples. Dashed, short dashed, and solid lines are guidelines for the eye. (c) Schematic energy band diagram of the CuPc/Co(Au) interface (for simplicity the average values are schematically plotted).

The energy band diagram [Fig. 2(c)] sketches the energy levels at the metal (Co or Au)–organic (CuPc) interface. On the left hand side, the bulk properties of the two metals are presented. To highlight the interface properties present at any M-O junction, the average energy values are given for the CuPc covered films (independent from metal substrate). The values on the right hand side (i.e., for CuPc films thicker than 10 nm) correspond to bulk-like CuPc properties.

The similarities and more interestingly the differences of the CuPc properties on the two metal substrates are discussed in the following. Previous thickness dependent UPS-IPS measurements of CuPc deposited on hydrogen passivated Si(111) (H-Si) show a "band bending"-like behavior at the S-O interface.[12] The band gap remains constant for all film thicknesses but the HOMO and LUMO shift parallel toward higher BEs with increasing film thickness. The overall value of the shift is (0.4 ± 0.1) eV.[12] As shown in Fig. 2(c), the M-O interface presents a "transport band gap opening." It saturates above 10 nm film thicknesses at a value of (2.2 ± 0.3) eV in agreement with the value for the transport band gap $E_t$ reported previously.[10–12] The transport band gap opening is the result of the image charge potential screening[21,22] for the very thin films.

The decrease of $E_t$ of molecular films close to the metal surface was reported by Tsiper *et al.*[23] In their UPS-IPS studies of perylenetetracarboxylic acid dianhydride (PTCDA) deposited on silver and scanning tunneling spectroscopy studies of PTCDA deposited on Au, they observed a reduction of $E_t$ by 400 meV (200meV in the occupied and 200 meV in the unoccupied electronic states regions) for monolayer coverage in comparison with thicker films. They explain this by the different polarizability of the metal substrate (at the interface) and the vacuum (at the film surface). It is also noted that the interface dipole (i.e., the work function of the substrate) has no influence on the reduction of $E_t$ . Knupfer and Paasch[21] even observed that molecules providing different interface dipoles (due to different IEs) on the same type of metallic substrate show the same amount of shifts of their HOMO levels. Moreover, MnPc deposited on Co foil shows a small band gap opening of 0.35 eV.[25] Finally, theoretical calculations of benzene physisorbed on graphite (0001) predict as well a strong renormalization of the electronic gap.[24]

In Fig. 2(a), $E_t$ is plotted with full and empty orange symbols for CuPc on Co and Au, respectively. It grows with film thickness from (1.4 ± 0.3) eV for 1 nm CuPc to the final value of (2.2 ± 0.3) eV for CuPc films >10 nm. The band gap opening is due to the shift of both the HOMO and the LUMO onsets away from the Fermi energy, shown in Fig. 2(b). These shifts can also be seen in IE and EA [Fig. 2(a)] even though these values are slightly influenced by changes in Φ.

The position of $E_v$ with regard to $E_F$ is plotted with triangles in Fig. 2(b) and it mirrors the behavior of Φ plotted in Fig. 2(a) with rectangles. Initially $E_v$ shows a large shift of ~0.8 eV toward $E_F$ for the 1 nm film followed by a smoother shift up to 1.0 eV and 1.2 eV on Co and Au substrates, respectively. These values saturate above 10 nm film thickness. The reason for the shift is the formation of an interface dipole Δ [plotted with green triangles in Fig. 2(a)]. On the Au substrate Δ has a value higher by 0.2 eV than for the Co substrate, i.e., 1.2 eV, in agreement with Peisert *et al.*[14] There, the authors investigated the interface between single crystalline and polycrystalline Au and CuPc molecules by UPS and XPS techniques. The formation of a 1.2 eV interface dipole and a shift of the HOMO toward higher BE by 0.25 eV was observed,[14] which is in good agreement with our observations. It was shown that CuPc on both single- and polycrystalline Au surfaces shows similar behavior for thicknesses larger than 1 nm, although the molecular growth is most probably quite different. The shift of the HOMO in the early stages was ascribed to a final state screening effect induced by the metallic substrate.[14] The complementary IPS measurements of our study provide a total shift of the LUMO onset position by 0.55 eV away from $E_F$. This provides the final bulk-like positions of the HOMO and the LUMO at 1.1 eV below and above $E_F$, respectively [see Fig. 2(c)].

The substrate of higher interest is the polycrystalline Co due to possible applications in spintronic devices like magnetic tunnel junctions.[7] CuPc behaves quite similarly on the Co and the Au substrate except for some essential differences. First, Co has a smaller work function, which induces a smaller interface dipole of 1.0 eV [green full triangles in Fig. 2(a)]. The second and more obvious difference is in the HOMO region [Fig. 2(b)]. The HOMO position is closer to $E_F$ by 0.2 eV which is highlighted in Fig. 2(b) by the eye guiding short dashed and solid lines. This difference is not induced by the interface dipole but more likely by charge transfer to the Co substrate. The onset of the HOMO of CuPc on Co saturates at larger thicknesses similarly as in the case of Au substrate. This will be discussed in much more detail in the following paragraphs after presenting the XPS results.

Figure 3 presents the core level spectra of CuPc deposited with increasing thickness on Au and Co foils plotted with empty circles and full rectangles, respectively. In Fig. 3(a), the spectra of Cu $2p_{3/2}$ are shown, which were fitted with a single asymmetric Voigt peak. The peak positions, marked by vertical gray lines, are different for the two samples and with increasing film thickness they both shift toward higher BE.

Figure 3(b) presents the N 1s spectra at different CuPc thicknesses deposited on Au and Co foils. The main peak at around 399 eV BE is the sum of contributions stemming from nitrogen atoms having different chemical environment, i.e., a lower BE contribution corresponds to the N atoms bound to the central metal ion ($N_M$) while a slightly higher BE contribution corresponds to the N atoms bound to the pyrrole rings ($N_P$). The source of the two components is schematically shown in Fig. 3(d). The peaks represented in Fig. 3(b) with red and blue patterns for $N_M$ and $N_P$ components, respectively, cannot be well resolved energetically. In the fit the line shapes and the intensities were coupled. The small feature at ~2 eV higher BE is the shake-up satellite peak of the main feature composed as well of two components, $SN_M$ and $SN_P$. The thickness evolution of the $N_M$ component is plotted with red hatched peak and its position is followed by the vertical gray lines. The peak shifts have similar behavior as in the case of the Cu $2p_{3/2}$ peak.

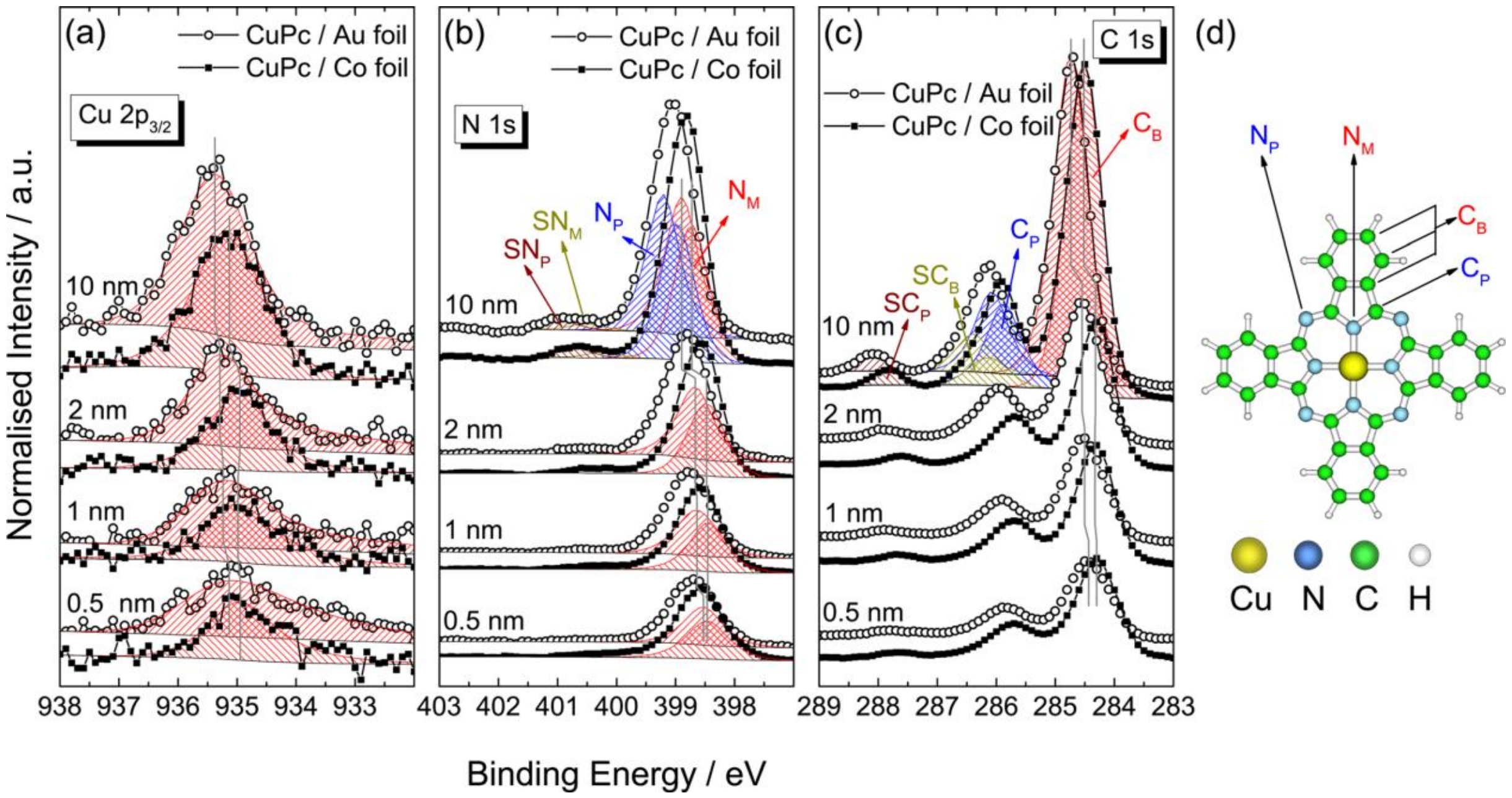

Figure 3. Core level spectra [Cu $2p_{3/2}$ (a), N 1 s (b), and C 1 s (c)] of CuPc films on Co (full rectangles) and Au (empty circles) substrates, respectively. For clarity, spectra at different thicknesses are shifted vertically with respect to each other with a slight additional shift between spectra of CuPc on the two substrates. $N_M$, $N_P$, $C_B$, and $C_P$ represent the N and C atoms with different chemical environments and $SN_M$, $SN_P$, $SC_B$, and $SC_P$ are their corresponding shake-up satellites. The gray vertical lines follow the evolution of the Cu $2p_{3/2}$, $N_M$, and $C_B$ features. (d) Schematic representation of the CuPc molecular structure and the labeling of different N and C species.

Between 284 eV and 288 eV BE the 1s peaks related to carbon atoms are visible [Fig. 3(c)]. The main feature at ~284.5 eV is assigned to C atoms from the benzene rings referred to as $C_B$ and sketched in Fig. 3(d). The second feature at ~286 eV is assigned to C atoms from the pyrrole rings, i.e., which are bonded to N, and is referred to as $C_P$. The small peaks labeled as $SC_B$ and $SC_P$ in Fig. 3(c) are the satellite peaks of $C_B$ and $C_P$, respectively. The sum of $C_P$ and $SC_P$ is three times smaller than the sum of $C_B$ and $SC_B$ as is expected from the molecule stoichiometry.29 All the C 1s peak shifts behave in a similar manner as the Cu $2p_{3/2}$ and N 1s features.

The shake-up satellites are the effect of the excitation of valence electrons into higher unfilled energy levels. The energy required for the transition leads to a reduction of the kinetic energy of the primary photoelectrons which makes them emerge at higher BEs in the spectra. The most likely transition is the one from HOMO to LUMO state; thus, the satellite to main peak distance is associated to be equal with the transport band gap ($E_t$).[29–31]

Figure 4(a) summarizes the peak positions versus the CuPc film thickness determined from the spectra of Figs. 3(a)–3(c). Note that the energy (vertical) scales are reversed for comparison with the HOMO onset positions in Fig. 2(b). Within the error bars all the core level peaks from CuPc on Co foil are at 0.2 eV lower BE than the ones for CuPc on Au foil. This is in good agreement with the HOMO positions within this thickness range. Because the UPS measurements are more surface sensitive than XPS for the 10 nm CuPc film they provide HOMO onset positions at the same energy on both substrates. The shifts of the core level peaks with increasing film thickness are between 0.20 eV and 0.25 eV and in good agreement with the 0.25 eV HOMO onset shifts.

The relative positions of the C 1s and N 1s shake-up satellites with regard to the main peaks are plotted in Fig. 4(b) by orange and green symbols, respectively, on both Au and Co substrates. No obvious difference is observed between satellites of CuPc deposited on the Au and Co substrates. The shake-up satellites of the benzene ($SC_B$) and pyrrole ($SC_P$) carbon features are relatively positioned to the main peak at 1.75 eV and 2.0 eV, respectively, and present no thickness dependency. On the other hand, the nitrogen related satellites ($SN_M$ and $SN_P$) have all the same split of 1.8 eV to the main peak at 10 nm thickness while for very thin films the satellites are positioned closer to the main feature (~1.6 eV) due to the band gap opening. The quantities determined for thick films are in good agreement with the literature data;[29,30] however, they are lower than $E_t$ = 2.2 eV determined by UPS-IPS. Moreover, some of the shake-up satellites present relative energy positions even lower than the optical band gap of CuPc ($E_{opt}$ = 1.76 eV).[10] This was observed for several molecules, see, e.g., Ref. 31. The distance between the shake-up satellite and the main peak is smaller compared to the transport band gap of the molecule due to final state screening of the core hole and to an optimized redistribution of charge within the molecule upon photoexcitation.[31]

All these results bring us to raise the questions: what is the reason for the peak shifts with film thickness and why is there a difference in the core level and valence band spectra between CuPc on Au and on Co substrates? Regarding the first question we already answered it in the previous paragraphs. The image charge potential not only reduces the transport band gap for very thin films it also influences the core levels, i.e., all spectra are shifted toward lower BEs as it was previously also observed for CuPc on Au.[14]

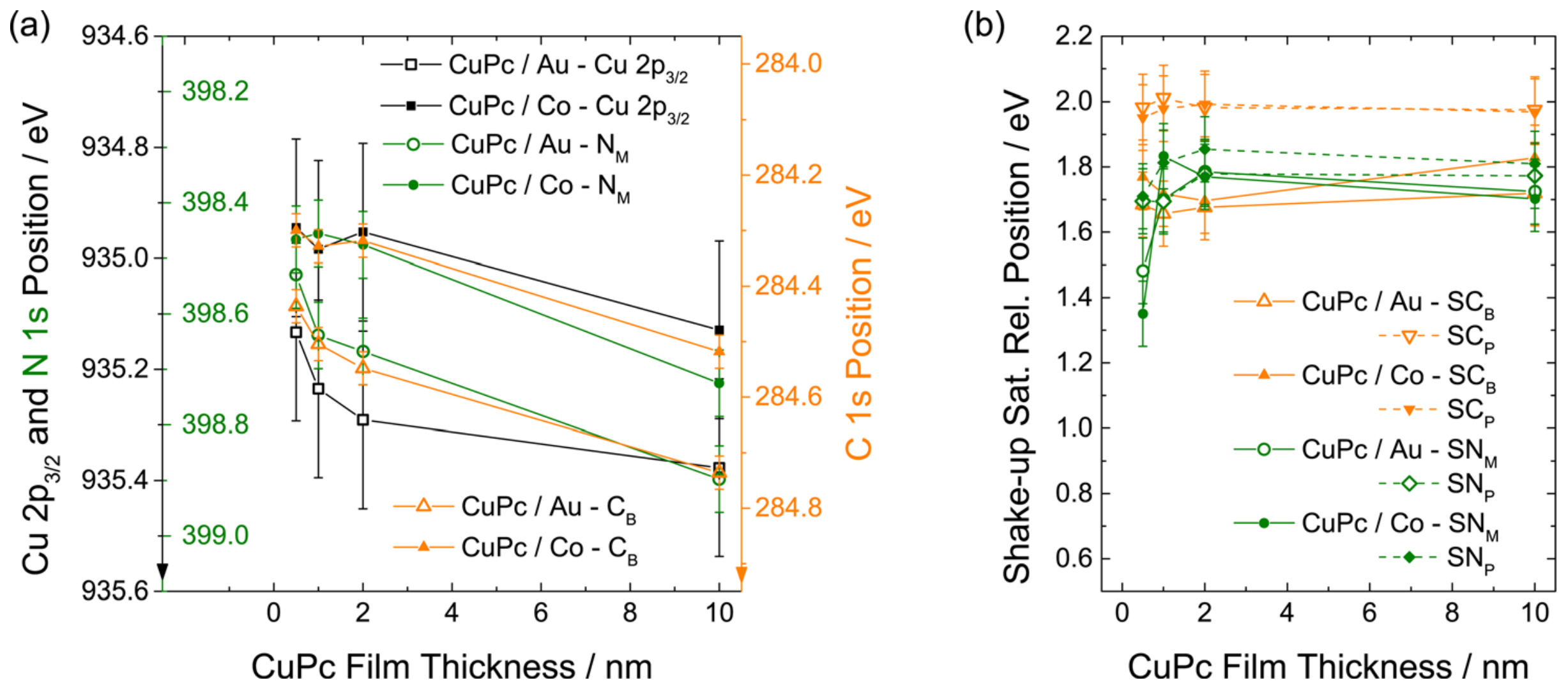

Figure 4. (a) Dependence of the core level peak positions as a function of film thickness. Data points of Cu $2p_{3/2}$ (rectangles), $N_M$ (circles), and $C_B$ (triangles), adopted after fitting the spectra in Fig. 3. Full and empty symbols are for CuPc on Co and Au substrates, respectively. (b) Energy positions of the shake-up satellites with regard to the main peaks as a function of the CuPc film thickness.

The second question can be addressed if we take a look at different metal–CuPc interfaces in literature. There are two geometries to determine the interface properties: evaporated CuPc on metal substrates (*bottom interface*) and evaporated metals on thick CuPc films (*top interface*). The first one excludes the possibility of metal atoms inter-diffusion into the organic layer. For the *bottom interface* a high reactivity of the Co surface was observed only for ultrathin sub-monolayer CuPc films deposited on Co(001).[5] At the *top interface* geometry where Co is evaporated on thick CuPc films,[8,9] the reactive interface between Co and CuPc is more obvious. Even though Co (Refs. 9 and 19) and Fe (Ref. 32) atoms do not diffuse as deep into the organic layer as Ag (Ref. 15) due to the higher reactivity with the molecules, they still react with many more molecules than in the *bottom interface* configuration. Co (Refs. 8, 9, and 19) and Fe (Ref. 32) atoms transfer charge to the CuPc molecules reducing the Cu atoms at the interface from Cu(II) to Cu(I). Our layer system is composed of rough polycrystalline Co foil on which an island-like growth mode of the molecules is expected. This is proven by the fact that the Fermi edge is visible up to a film thickness of 6 nm in the UP and IP spectra [Figs. 2(c) and 2(g)]. Due to the ill-defined interface of the samples, no monovalent Cu(I) state could be observed in the Cu $2p_{3/2}$ core level spectra at low thicknesses (above one monolayer coverage), as was shown by Lach *et al.*[5] (sub-monolayer coverage). A full picture of the metal–organic–metal system Co/pentacene/Co shows that the two interfaces are not really symmetric and should not be considered identical.[19]

Considering for comparison the Ag/CuPc interface, in the *top interface* configuration Ag shows no chemical reaction and high diffusion into the CuPc layer.[15] With increasing Ag thickness on CuPc the HOMO shifts by 0.2 eV closer to $E_F$,[15] which is a sign of charge transfer from CuPc into the upper Ag layer, i.e., a p-type doping of the underlying CuPc film. Thus, this finding is similar to the CuPc/Co interface studied in this work. As a *bottom interface*, the partially fluorinated CuPc molecule ($CuPcF_4$) deposited on Ag substrate was taken into consideration.[33] However, in this second case, the interface study shows a net charge transfer from Ag to the more electronegative $CuPcF_4$ molecules. According to Schwieger *et al.* the $CuPcF_4$/Ag interface has similar properties with the *n-type* doping by K intercalation.[33] Comparing the valence band spectra of $CuPcF_4$ films with a thickness of 2 nm deposited on a silver substrate the HOMO onset position is at 0.4 eV higher BE than the one deposited on gold substrate.[33] Since in our case the HOMO onset of the CuPc on Co is at 0.2 eV lower BE than the one deposited on Au we come to the conclusion that at the CuPc/Co interface there is a net charge transfer from the CuPc molecules to the Co substrate, hence a kind of *p-type* doping of the molecular film.

## IV. Summary and Conclusions

Metal–organic interfaces play a major role for the newly increasing field of organic spintronics. CuPc is a highly stable organic molecule with a high spin injection efficiency[3] and long spin relaxation time and cobalt is an intensively studied metal contact for spin injection into organic materials.[3–9] However, knowledge about the bulk spin transport properties of the materials is not sufficient for the design of efficient spin valve devices. In organic spintronics, both the electronic and spin properties at the interface, the so called "spinterface,"[34] play a major role to control the spin flow. The HOMO and LUMO levels positions of the molecules at the M-O interface, i.e., the energy band alignment, is different compared to the bulk materials which influences the charge and spin flow in such devices. Combined UPS and IPS measurements on organic thin films function of thickness provide valuable information on the energy band evolution.

The transport band gap of CuPc films above 10 nm was determined to be (2.2 ± 0.3) eV in good agreement with literature.[10–12] For thinner films it decreases down to a value of (1.4 ± 0.3) eV for 1 nm CuPc film thickness on both Co and Au substrates. This closing of the transport band gap for very thin films is induced by the image charge potential[21–23] mirrored in the metal substrates. Due to charge screening the energy levels of the molecular orbitals are drawn closer to the Fermi level. As a consequence, both the HOMO and LUMO onset positions shift closer to the Fermi level by 0.25 eV and 0.55 eV, respectively, when going from thick (20 nm) to very thin (1 nm) CuPc films. The interface dipoles at the CuPc/Co and CuPc/Au interfaces were determined to be (1.0 ± 0.1) eV and (1.2 ± 0.1) eV, respectively. These quantities depend on the work functions of the two metal substrates.

In addition, the HOMO onset positions and the core levels positions at the CuPc/Co interface have an offset of 0.2 eV toward lower binding energy as compared to the CuPc/Au interface. Even though there should be electron donation from the Co atoms of the substrate to the central Cu atom of the molecules reducing it from Cu(II) to Cu(I),[8,9] a net charge is transferred from the macrocycle of the Pc molecule to the Co substrate. This introduces a shift in the electronic levels toward lower BEs, i.e., a shift of the Fermi edge, as would be expected at a *p-type* doping.

Consequently, tailoring the interface chemical environment highly influences not only the spin injection capabilities[5] but also the energy level alignment at the metal–organic interface, which is a crucial element in both electronic and spintronic devices.

## Acknowledgments

This work was supported by the Deutsche Forschungsgemeinschaft (DFG) Research Unit FOR 1154 "Towards Molecular Spintronics". The authors thank HZB for the allocation of synchrotron radiation beamtime. The authors are also grateful to the staff of the Russian German Laboratory for their great support at the beamline.